\documentclass[doublecol]{epl2} 

\title{Virial theorem  for rotating self-gravitating Brownian particles \\
and two-dimensional point vortices}
\shorttitle{Virial theorem for rotating self-gravitating Brownian particles and 2D point vortices} 

\author{Pierre-Henri Chavanis\inst{1,2}}
\shortauthor{Pierre-Henri Chavanis}

\institute{                    
  \inst{1}Universit\'e de Toulouse, UPS,  Laboratoire de Physique Th\'eorique (IRSAMC), F-31062 Toulouse, France \\
  \inst{2} CNRS, Laboratoire de Physique Th\'eorique (IRSAMC), F-31062 Toulouse, France
}
 
\abstract{We derive the proper form of Virial theorem for a system of rotating self-gravitating Brownian particles. We show that, in the two-dimensional case, it takes a very simple form that can be used to obtain general results about the dynamics of the system without being required to solve the Smoluchowski-Poisson system explicitly. We also develop the analogy between self-gravitating systems and two-dimensional point vortices and derive a Virial-like relation for the vortex system.}

\begin{document}

\maketitle

\section{\label{intro}Introduction}

The theory of Brownian motion is a fundamental topic in physics since
the pioneering works of Einstein and Smoluchowski
\cite{risken}. Brownian theory has applications in many other area
including chemistry, biology and finance. The study of Brownian
particles in interactions was first developed in connection to
colloidal suspensions, inhomogeneous fluids, and supercooled liquids
exhibiting glassy features \cite{evans,dean,cugliandolo}. In that
case, the interaction is short-ranged. The study of Brownian particles
with long-range interactions is more recent. For example, the case of
self-gravitating Brownian particles has been studied by Chavanis and
Sire in a series of papers (see \cite{cs} and references therein). For
this system, the mean field approximation is exact in a proper
thermodynamic limit $N\rightarrow +\infty$ with $\eta=\beta
GMm/R^{d-2}$ fixed (where $d$ is the dimension of space) and it leads
to the Smoluchowski-Poisson system. The Smoluchowski-Poisson (SP)
system displays many formal analogies with the Keller-Segel (KS) model
\cite{ks} of chemotaxis for bacterial populations  in biology 
(see \cite{mass} for a description of this analogy). In recent years,
the SP system and the KS model have been the object of intense
research from physicists (see, e.g.,
\cite{cs} and references therein) and applied mathematicians (see, e.g., 
\cite{kav} and references therein).

For self-gravitating systems, the Virial theorem plays a very
important role \cite{bt}. For a two-dimensional (2D) self-gravitating
Brownian gas, it has been shown in \cite{cs,mass,exact} that the
Virial theorem takes a very simple {\it closed} form. In this Letter,
we shall extend this relation to the case of a rotating
self-gravitating Brownian gas. So far, there is no result on the
rotating case. From the Virial theorem, we can deduce general
properties of the rotating SP system without being required to solve
these equations explicitly.  In this Letter, we focus on the basic
results and make simplifying assumptions: (i) mean field
approximation; (ii) single species system; (iii) overdamped
dynamics. The general case, relaxing these assumptions, will be
considered in a companion paper (in preparation). We shall also
develop the analogy between self-gravitating systems and 2D point
vortices initiated in \cite{houches} and derive a Virial-like relation
for the vortex system.

\section{Virial theorem for a rotating self-gravitating Brownian gas}

\subsection{The Virial theorem}

We consider a single species system of self-gravitating Brownian particles with individual mass $m$ rotating with angular velocity $\Omega$. We work in the rotating frame. In the mean field approximation, and in the strong friction (overdamped) limit $\xi\rightarrow +\infty$,  the dynamical evolution of the density $\rho({\bf r},t)$ is described by the Smoluchowski-Poisson system \cite{cs,cll}:
\begin{eqnarray}
\label{e1}
\frac{\partial\rho}{\partial t}=\nabla \cdot \left \lbrack \frac{1}{\xi}\left (\nabla p+\rho\nabla\Phi_{eff}\right )\right\rbrack,
\end{eqnarray}
\begin{eqnarray}
\label{e2}
\Delta\Phi=S_d G\rho,
\end{eqnarray}
where
\begin{eqnarray}
\label{e3}
\Phi_{eff}({\bf r},t)=\Phi({\bf r},t)-\frac{(\Omega\times {\bf r})^{2}}{2},
\end{eqnarray}
is the effective gravitational potential accounting for inertial
forces and $S_d$ is the surface of a unit sphere in $d$ dimensions (in
this paper, $d=3$ or $d=2$ when $\Omega\neq 0$ and $d$ is arbitrary
when $\Omega=0$). For the sake of generality, we consider the
generalized Smoluchowski-Poisson (GSP) system \cite{cll} with an
arbitrary barotropic equation of state $p=p(\rho)$ but, for ordinary
self-gravitating Brownian particles, the equation of state is the
isothermal one
\begin{eqnarray}
\label{e4}
p({\bf r},t)=\rho({\bf r},t)\frac{k_B T}{m}.
\end{eqnarray}
Since the temperature is fixed, we are describing a dissipative gas in the canonical ensemble.

The Lyapunov functional associated with the isothermal SP system is
the free energy $F=E_J-TS_B$ where $E_{J}=\frac{1}{2}\int\rho\Phi\,
d{\bf r}-\frac{1}{2}\int \rho (\Omega\times {\bf r})^{2}\, d{\bf r}$
is the Jacobi energy and $S_B=-k_{B}\int \frac{\rho}{m}\ln
\frac{\rho}{m} \, d{\bf r}$ is the configurational Boltzmann
entropy. For the GSP system, the Lyapunov functional is the
generalized free energy
$F=E_J+\int\rho\int^{\rho}\frac{p(\rho')}{\rho^{'2}}\, d\rho'\, d{\bf
r}$ \cite{cll}. Using Eqs. (\ref{e1})-(\ref{e2}), we easily derive the
canonical $H$-theorem
\begin{eqnarray}
\label{e5}
\dot F=-\int \frac{1}{\xi\rho}\left (\nabla p+\rho\nabla\Phi_{eff}\right )^2\, d{\bf r} \le 0.
\end{eqnarray}
At equilibrium, $\dot F=0$, implying
\begin{eqnarray}
\label{e6}
\nabla p+\rho\nabla\Phi_{eff}={\bf 0},
\end{eqnarray}
which is the condition of hydrostatic equilibrium in the rotating frame. For the isothermal equation of state (\ref{e4}), we obtain, after integration, the mean field Boltzmann distribution
\begin{eqnarray}
\label{e7}
\rho({\bf r})=A e^{-\beta m\Phi_{eff}({\bf r})},
\end{eqnarray}
where $\beta=1/(k_B T)$. It can be shown that this distribution is
dynamically stable with respect to the SP system if and only if (iff)
it is a (local) minimum of free energy at fixed mass. In that case, it
represents the statistical equilibrium state in the canonical ensemble
\cite{cs}. In the microcanonical ensemble, the statistical
equilibrium state is a (local) maximum of entropy at fixed mass,
energy and angular momentum \cite{rieutord}. The critical points
(canceling the first order variations of entropy) are also given by
the mean field Boltzmann distribution (\ref{e7}). Therefore, all the
results obtained in the following at equilibrium, that exclusively use
the form of Eq. (\ref{e7}), are valid both in the canonical and in the
microcanonical ensembles \footnote{The inequivalence between the
microcanonical and the canonical ensembles appears only when we
consider the thermodynamical stability of the system which is related
to the sign of the second order variations of entropy or free energy
\cite{paddy,aa}.} .

The equilibrium scalar Virial theorem for self-gravitating Brownian particles can be derived as follows.
Taking the scalar product of Eq. (\ref{e6}) with ${\bf r}$, and using the identity $\frac{1}{2}\nabla ((\Omega\times {\bf r})^2)=-\Omega\times (\Omega\times {\bf r})=\Omega^2{\bf r}-(\Omega\cdot {\bf r})\Omega$ we obtain
\begin{eqnarray}
\label{e8}
\int \nabla p\cdot {\bf r}\, d{\bf r}-W_{ii}-I\Omega^{2}+I_{ik}\Omega_{k}\Omega_{i}={0},
\end{eqnarray}
where we have introduced  the tensor of inertia $I_{ij}=\int\rho x_i x_j\, d{\bf r}$ (its trace is the moment of
inertia $I=\int\rho r^2\, d{\bf r}$) and the Virial $W_{ii}=-\int \rho {\bf r}\cdot \nabla\Phi \, d{\bf r}$. For $d\neq 2$,  $W_{ii}=(d-2)W$ where $W=\frac{1}{2}\int \rho\Phi\, d{\bf r}$ is the potential energy and for $d=2$, $W_{ii}=-\frac{GM^2}{2}$ \cite{cs}. Integrating the first term by parts, we obtain
\begin{eqnarray}
\label{e9}
d\int p\, d{\bf r}+W_{ii}+I\Omega^{2}-I_{ik}\Omega_{k}\Omega_{i}=dPV,
\end{eqnarray}
where we have introduced the average pressure of the system on the
boundary of the domain $P=\frac{1}{dV}\oint p {\bf r}\cdot d{\bf S}$
(if the pressure is uniform on the boundary: $p({\bf r})=p_b$, then
$P=p_b$). For an unlimited system, $PV\rightarrow 0$, provided that
the pressure decreases sufficiently rapidly with the distance.  For
the isothermal equation of state (\ref{e4}), we obtain
\begin{eqnarray}
\label{e10}
dNk_B T+W_{ii}+I\Omega^{2}-I_{ik}\Omega_{k}\Omega_{i}=dPV.
\end{eqnarray}

The out-of-equilibrium Virial theorem is obtained by taking the time derivative of the moment of inertia and using the GSP system Eqs. (\ref{e1})-(\ref{e2}). After straightforward algebra, we obtain
\begin{eqnarray}
\label{e11}
\frac{1}{2}\xi {\dot I}=d\int p\, d{\bf r}+W_{ii}+I\Omega^{2}
-I_{ik}\Omega_{k}\Omega_{i}
-dPV.
\end{eqnarray}
For an (isothermal) Brownian gas
\begin{eqnarray}
\label{e12}
\frac{1}{2}\xi {\dot I}=dNk_{B}T+W_{ii}+I\Omega^{2}
-I_{ik}\Omega_{k}\Omega_{i}
-dPV.
\end{eqnarray}

\subsection{The two-dimensional case}

For a 2D system of self-gravitating Brownian particles, the Virial theorem takes a particularly simple form. In that case, $\Omega=\Omega{\bf z}$, $\Phi_{eff}=\Phi-\frac{1}{2}\Omega^{2}r^{2}$ and $\nabla (\Omega r^2)=2\Omega {\bf r}$. The scalar Virial theorem (\ref{e11}) becomes
\begin{eqnarray}
\label{e13}
\frac{1}{2}\xi {\dot I}=2\int p\, d{\bf r}-\frac{GM^2}{2}+I\Omega^{2}
-2PV.
\end{eqnarray}
For the isothermal equation of state (\ref{e4}), we obtain
\begin{eqnarray}
\label{e14}
\frac{1}{2}\xi {\dot I}=2Nk_B T-\frac{GM^2}{2}+I\Omega^{2}
-2PV.
\end{eqnarray}
Introducing the  mean field critical temperature
\begin{eqnarray}
\label{e15}
k_{B}T_{c}=\frac{GNm^2}{4},
\end{eqnarray}
it can be rewritten
\begin{eqnarray}
\label{e16}
\frac{1}{2}\xi {\dot I}=2Nk_{B}(T-T_{c})+I\Omega^{2}
-2PV.
\end{eqnarray}
This relation is the main result of this Letter. It extends the Virial theorem given in \cite{cs,mass,exact}  for non-rotating systems ($\Omega=0$). At equilibrium,
\begin{eqnarray}
\label{e17}
PV=Nk_{B}(T-T_{c})+\frac{1}{2}I\Omega^{2}.
\end{eqnarray}
This can be viewed as the global equation of state for a 2D rotating self-gravitating isothermal gas. As explained previously, it is valid both in the canonical and in the microcanonical ensembles.  For a non-rotating system ($\Omega=0$), it reduces to 
\begin{eqnarray}
\label{e18}
PV=Nk_{B}(T-T_{c}).
\end{eqnarray}
This equation of state has been  previously derived by Salzberg \cite{salzberg}, Katz \& Lynden-Bell \cite{klb} and Chavanis \cite{mass,exact}, using different methods.
Since $P\ge 0$, this relation shows that equilibrium states can possibly  exist only for $T\ge T_{c}$. In an infinite domain ($P=0$), they can possibly exist only at the critical temperature $T=T_{c}$ (see \cite{cs,mass} for more details and explicit solutions). For a rotating system in an infinite domain ($P=0$),  the equilibrium Virial theorem reduces to
\begin{eqnarray}
\label{e19}
\frac{1}{2}I\Omega^{2}=Nk_{B}(T_{c}-T).
\end{eqnarray}
Since $I\Omega^{2}\ge 0$, this relation shows that equilibrium states can possibly exist only for $T\le T_{c}$.

In an infinite domain ($P=0$), the Virial theorem (\ref{e16}) for a rotating self-gravitating Brownian gas reduces to
\begin{eqnarray}
\label{e20}
\frac{1}{2}\xi {\dot I}-\Omega^{2}I=2Nk_{B}(T-T_{c}).
\end{eqnarray}
It is worth noticing that this equation is closed, which is not the
case for the Virial theorem of Hamiltonian systems \cite{bt}. This
simplification for Brownian systems is due to the strong friction
limit and to the two-dimensional assumption. Solving this equation, we
find that the moment of inertia evolves like
\begin{eqnarray}
\label{e21}
I(t)=\left\lbrack I(0)-\frac{2Nk_{B}}{\Omega^{2}}(T_{c}-T)\right\rbrack e^{\frac{2\Omega^{2}}{\xi}t}\nonumber\\
+\frac{2Nk_{B}}{\Omega^{2}}(T_{c}-T).
\end{eqnarray}
This relation shows that the system will experience collapse or evaporation depending on the sign of the quantity in bracket. Let us introduce the new critical temperature
\begin{eqnarray}
\label{e22}
T_{0}=T_{c}-\frac{I(0)\Omega^{2}}{2Nk_{B}}.
\end{eqnarray}
This critical temperature depends on the initial value of the moment of inertia and on the angular velocity. It can be written
\begin{eqnarray}
\label{e23}
T_{0}(\Omega)=T_{c}\left (1-\frac{\Omega^{2}}{\Omega_{0}^{2}}\right ),
\end{eqnarray}
with
\begin{eqnarray}
\label{e24}
\Omega_{0}=\left\lbrack \frac{GM^{2}}{2I(0)}\right \rbrack^{1/2}.
\end{eqnarray}
We note that $T_0\le T_c$. In terms of the critical temperature (\ref{e23}), the evolution of the moment of inertia is given by
\begin{eqnarray}
\label{e25}
I(t)=\frac{2Nk_{B}}{\Omega^{2}}\left\lbrack (T_{c}-T)+(T-T_{0}) e^{\frac{2\Omega^{2}}{\xi}t}\right\rbrack.
\end{eqnarray}
For small times, we have
\begin{eqnarray}
\label{e26}
I(t)\simeq I(0)+\frac{4Nk_{B}}{\xi}(T-T_{0})t, \qquad (t\rightarrow 0).
\end{eqnarray}

For $T>T_{0}$, the moment of inertia $I(t)$ increases  and tends
to $+\infty$ as $t\rightarrow +\infty$. Since $I(t)=M\langle r^2\rangle(t)$, this corresponds to an evaporation process. For $T<T_{0}$, the moment of inertia $I(t)$ decreases and becomes zero in a finite time
\begin{eqnarray}
\label{e27}
t_{end}=\frac{\xi}{2\Omega^{2}}\ln\left (\frac{T_{c}-T}{T_{0}-T}\right ).
\end{eqnarray}
This corresponds to a finite time {collapse}. At $t=t_{end}$, the moment of inertia $I=0$, implying that the system forms a Dirac peak containing the whole mass $M$.  For $T=T_{0}$, the moment of inertia remains constant for all times: 
\begin{eqnarray}
\label{e28}
I(t)=\frac{2N}{\Omega^{2}}k_{B}(T_{c}-T_0)=I(0).
\end{eqnarray}

For a non rotating infinite system ($\Omega=P=0$), we recover the Virial theorem obtained in \cite{cs,mass,exact}:
\begin{eqnarray}
\label{e16n}
\frac{1}{2}\xi {\dot I}=2Nk_{B}(T-T_{c}).
\end{eqnarray}
We can define an effective diffusion coefficient
\begin{eqnarray}
\label{e16nn}
D_{eff}(T)=\frac{k_{B}T}{\xi m}\left (1-\frac{T_{c}}{T}\right ),
\end{eqnarray}
so that $\langle r^2\rangle=4D_{eff}(T)t+\langle r^2\rangle_0$.
Eq. (\ref{e16n}) can be integrated into 
\begin{eqnarray}
\label{e29}
I(t)=I(0)+\frac{4Nk_{B}}{\xi}(T-T_{c})t.
\end{eqnarray}
For $T>T_{c}$, the system evaporates and for $T<T_{c}$, it experiences  gravitational collapse in a finite time
\begin{eqnarray}
\label{e30}
t_{end}=\frac{\xi I(0)}{4Nk_{B}(T_{c}-T)}.
\end{eqnarray}
At $t=t_{end}$, the moment of inertia $I=0$, implying that the system forms a Dirac peak containing the whole mass $M$. For $T=T_c$, the moment of inertia is conserved (see \cite{cs,mass} for more details and explicit solutions).

The above results are very general and do not depend whether the system is axisymmetric or not. It is interesting that we can obtain these qualitative behaviors without being required to solve the SP system (\ref{e1})-(\ref{e2}).

\section{Virial theorem for 2D point vortices}

We now consider a single species system of point vortices in two-dimensional hydrodynamics. It has been shown in \cite{bv} that the Gibbs canonical distribution is reproduced by the statistical equilibrium  state of a system of Brownian point vortices whose $N$-body dynamics is defined in terms of stochastic equations (instead of deterministic equations for usual point vortices \cite{newton}). This is the counterpart of the system of self-gravitating Brownian particles considered previously.  In the mean field approximation, the evolution of the smooth vorticity field $\omega({\bf r},t)$ is governed by the Fokker-Planck equation \cite{bv}:
\begin{eqnarray}
\label{e31}
\frac{\partial\omega}{\partial t}=\nabla \cdot \left \lbrack D\left (\nabla \omega+\beta\gamma\omega\nabla\psi_{eff}\right )\right\rbrack,
\end{eqnarray}
\begin{eqnarray}
\label{e32}
-\Delta\psi=\omega,
\end{eqnarray}
where $\gamma$ is the circulation of a point vortex, $\beta=1/(k_B T)$ is the inverse temperature (which can be positive or negative \cite{onsager}) and  $\psi_{eff}({\bf r},t)=\psi({\bf r},t)+\frac{1}{2}\Omega r^2$ is the relative stream function in the rotating frame.
The Lyapunov functional associated to this equation is the Massieu function (free energy) $J=S_B-\beta E_{eff}$ where $S_B=-\int \frac{\omega}{\gamma}\ln  \frac{\omega}{\gamma}\, d{\bf r}$ is the Boltzmann entropy of point vortices and
$E_{eff}=\frac{1}{2}\int \omega\psi \, d{\bf r}+\frac{1}{2}\Omega\int \omega r^2\, d{\bf r}$ is the effective energy. Using Eqs. (\ref{e31})-(\ref{e32}), we easily derive the canonical $H$-theorem
\begin{eqnarray}
\label{e33}
\dot J=\int \frac{D}{\gamma \omega}\left (\nabla \omega+\beta\gamma\omega\nabla\psi_{eff}\right )^2\, d{\bf r}\ge 0.
\end{eqnarray}
At equilibrium, $\dot J=0$, implying
\begin{eqnarray}
\label{e34}
\frac{k_B T}{\gamma}\nabla \omega+\omega\nabla\psi_{eff}={\bf 0},
\end{eqnarray}
which is formally similar to the condition of hydrostatic equilibrium
(\ref{e6}) for self-gravitating systems with a ``pressure"
$p=\frac{k_B T}{\gamma}\omega$ \cite{houches}. Integrating
Eq. (\ref{e34}), we obtain the mean field Boltzmann distribution
\begin{eqnarray}
\label{e35}
\omega({\bf r})=Ae^{-\beta \gamma\psi_{eff}({\bf r})}.
\end{eqnarray}
This distribution is dynamically stable with respect to the
Fokker-Planck equation (\ref{e31})-(\ref{e32}) iff it is a maximum of
Massieu function at fixed circulation. In that case, it represents the
statistical equilibrium state in the canonical ensemble \cite{bv}. In
the microcanonical ensemble, the statistical equilibrium state
maximizes the entropy at fixed circulation, energy and angular
momentum \cite{jm,cl}. The critical points (canceling the first order
variations of entropy) are also given by the mean field Boltzmann
distribution (\ref{e35}). Therefore, all the results obtained in the
following at equilibrium, that exclusively use the form of
Eq. (\ref{e35}) are valid both in the canonical and in the
microcanonical ensembles.

We now derive the equilibrium Virial theorem of point vortices. Taking the scalar product of Eq. (\ref{e34}) with ${\bf r}$ and integrating over the whole plane, we obtain
\begin{eqnarray}
\label{e36}
\frac{k_B T}{\gamma}\int \nabla \omega\cdot {\bf r}\, d{\bf r}+{\cal V}+\Omega L={0},
\end{eqnarray}
where we have introduced the angular momentum $L=\int\omega r^2\, d{\bf r}$ (similar to the moment of inertia for material particles) and defined the ``Virial of point vortices" ${\cal V}=\int \omega {\bf r}\cdot \nabla\psi\, d{\bf r}$ by analogy with the Virial of the gravitational force. Using the expression of the stream function $\psi({\bf r})=-\frac{1}{2\pi}\int \omega({\bf r}')\ln|{\bf r}-{\bf r}'|\, d{\bf r}'$ in an infinite domain, we easily find that ${\cal V}=-\frac{\Gamma^2}{4\pi}$. Integrating the first term of Eq. (\ref{e36})  by parts, we obtain the equilibrium Virial theorem
\begin{eqnarray}
\label{e37}
-2Nk_BT-\frac{\Gamma^2}{4\pi}+\Omega L=0.
\end{eqnarray}
Introducing the mean field critical temperature
\begin{eqnarray}
\label{e38}
k_{B}T_{c}=-\frac{N\gamma^2}{8\pi},
\end{eqnarray}
it can be rewritten
\begin{eqnarray}
\label{e39}
\frac{1}{2}\Omega L=Nk_B(T-T_c).
\end{eqnarray}
As explained previously, this relation is valid both in the canonical
and in the microcanonical ensembles. This relation was previously
derived by Pointin \& Lundgren \cite{pl} by a very different
method. We show here that it can be interpreted as the Virial theorem
for a point vortex system.  For $L=0$ or $\Omega=0$, we deduce that an
equilibrium state can possibly exist in an infinite domain only at the
critical temperature $T=T_c$.

The out-of-equilibrium Virial theorem for Brownian point vortices is obtained by taking the time derivative of the angular momentum and using the mean field Fokker-Planck equation (\ref{e31})-(\ref{e32}). After straightforward algebra, we obtain
\begin{eqnarray}
\label{e40}
\frac{k_B T}{4D\gamma} \dot{L}+\frac{1}{2}\Omega L=Nk_{B}(T-T_{c}).
\end{eqnarray}
Solving this equation, we find that the angular momentum evolves like
\begin{eqnarray}
\label{e41}
L(t)=\left\lbrack L(0)-\frac{2Nk_{B}}{\Omega}(T-T_c)\right\rbrack e^{-\frac{2D\gamma\Omega t}{k_B T}}\nonumber\\
+\frac{2Nk_{B}}{\Omega}(T-T_{c}).
\end{eqnarray}
Let us introduce the new critical temperature
\begin{eqnarray}
\label{e42}
T_{0}=T_{c}+\frac{\Omega L(0)}{2Nk_{B}}.
\end{eqnarray}
This critical temperature depends on the initial value of the angular momentum and on the angular velocity. It can be written
\begin{eqnarray}
\label{e43}
T_{0}(\Omega)=T_{c}\left (1-\frac{\Omega}{\Omega_{0}}\right ),
\end{eqnarray}
with
\begin{eqnarray}
\label{e44}
\Omega_{0}=\frac{N^{2}\gamma^2}{4\pi L(0)}.
\end{eqnarray}
In terms of the critical temperature (\ref{e42}), the evolution of the angular momentum is given by
\begin{eqnarray}
\label{e45}
L(t)=\frac{2Nk_{B}}{\Omega}\left\lbrack (T-T_{c})+(T_{0}-T) e^{-\frac{2D\gamma\Omega}{k_B T}t}\right\rbrack.
\end{eqnarray}
For small times, we have
\begin{eqnarray}
\label{e46}
L(t)\simeq L(0)+\frac{4D\gamma N}{T}(T-T_{0})t, \qquad (t\rightarrow 0).
\end{eqnarray}
In the following, we take $\gamma>0$ (the case $\gamma<0$ can be treated symmetrically).

(i) Let us first assume  $\Omega>0$. In that case, $T_0>T_c$. If $T>0$, then $L(t)\rightarrow \frac{2Nk_{B}}{\Omega}(T-T_{c})$ for $t\rightarrow +\infty$ (it increases for $T>T_0$ and it decreases for $T<T_0$). The system is expected to reach an equilibrium state. If $T<0$, we need to consider three cases: (a) if $T<T_0$, the angular momentum $L(t)$ increases  and tends
to $+\infty$ as $t\rightarrow +\infty$. Since $L(t)=\Gamma\langle r^2\rangle(t)$, the system of Brownian vortices ``evaporates". (b) If $T>T_0$, the angular momentum $L(t)$ decreases and becomes zero in a finite time
\begin{eqnarray}
\label{e47}
t_{end}=-\frac{k_B T}{2D\gamma \Omega}\ln\left (\frac{T-T_{c}}{T-T_{0}}\right ).
\end{eqnarray}
This corresponds to a finite time {collapse}. At $t=t_{end}$, the angular momentum $L=0$, implying that the system forms a Dirac peak containing all the point vortices. (c) If $T=T_{0}$, the angular momentum remains constant for all times:
\begin{eqnarray}
\label{e48}
L(t)=\frac{2N}{\Omega}k_{B}(T_{0}-T_c)=L(0).
\end{eqnarray}

(ii) Let us now assume  $\Omega<0$. In that case, $T_0< T_c$. If $T>0$, the angular momentum $L(t)$ increases  and tends
to $+\infty$ as $t\rightarrow +\infty$, leading to an evaporation process. If $T<0$, we need to consider three cases: (a) if $T<T_c$, then $L(t)\rightarrow \frac{2Nk_{B}}{\Omega}(T-T_{c})$ for $t\rightarrow +\infty$ (it increases for $T<T_0$ and it decreases for $T>T_0$). The system is expected to reach an equilibrium state. (b) If $T>T_c$, the angular momentum $L(t)$ decreases and becomes zero in a finite time given by Eq. (\ref{e47}). This corresponds to a finite time {collapse}. At $t=t_{end}$, the angular momentum $L=0$, implying that the system forms a Dirac peak containing all the point vortices. (c) If $T=T_c$, then
\begin{eqnarray}
\label{e49}
L(t)=L(0)e^{-\frac{2D\gamma\Omega}{k_B T_c}t}.
\end{eqnarray}
The angular momentum decreases and tends to zero for $t\rightarrow +\infty$, implying that the system forms a Dirac peak containing all the point vortices in infinite time.

For $\Omega=0$, the Virial theorem (\ref{e40}) reduces to
\begin{eqnarray}
\label{e40n}
\dot{L}=\frac{4D\Gamma}{T}(T-T_{c}).
\end{eqnarray}
We can define an effective diffusion coefficient
\begin{eqnarray}
\label{e40nn}
D_{eff}(T)=D\left (1-\frac{T_{c}}{T}\right ),
\end{eqnarray}
so that $\langle r^2\rangle=4D_{eff}(T)t+\langle r^2\rangle_0$.
Eq. (\ref{e40n}) can be integrated into
\begin{eqnarray}
\label{e50}
L(t)=L(0)+\frac{4DN\gamma}{T}(T-T_{c})t.
\end{eqnarray}
For $T>0$ and for $T<T_{c}$ (i.e. $\beta>\beta_c$), the angular
momentum increases and tends to $+\infty$ for $t\rightarrow +\infty$
so the system evaporates. For $T_{c}<T<0$ (i.e. $\beta<\beta_c$), the
angular momentum decreases and becomes zero in a finite time
\begin{eqnarray}
\label{e51}
t_{end}=\frac{-L(0)T}{4DN\gamma(T-T_{c})}.
\end{eqnarray}
At $t=t_{end}$, the angular momentum  $L=0$, implying that the system forms a Dirac peak containing all the point vortices. For $T=T_c$, the angular momentum is conserved.

\section{Conclusion}

In this Letter, we have shown that the Virial theorems for 2D self-gravitating Brownian particles and point vortices take a very simple form. While the out-of-equilibrium Virial theorems Eqs. (\ref{e12}), (\ref{e13}), (\ref{e16}) and  (\ref{e40}) are  only valid in the canonical ensemble (for Brownian systems), the equilibrium Virial theorems Eqs. (\ref{e10}), (\ref{e17}) and  (\ref{e39}) are valid both in canonical and microcanonical  ensembles.
To make the discussion straightforward, and emphasize the basic results, we have considered a single species system and made the mean field approximation. We shall relax these assumptions in a companion paper. Interestingly, we find that the exact Virial theorems keep the same form as the mean field Virial theorems except that the mean field critical temperatures are replaced by the {\it exact} critical temperatures \cite{exact}:
\begin{eqnarray}
\label{e52}
k_{B}T_{c}=\frac{G}{4N}\sum_{\alpha=1}^{N}\sum_{\alpha'\neq \alpha}m_{\alpha}m_{\alpha'},
\end{eqnarray}
\begin{eqnarray}
\label{e53}
 k_{B}T_{c}=-\frac{1}{8\pi N}\sum_{\alpha=1}^{N}\sum_{\alpha'\neq \alpha}\gamma_{\alpha}\gamma_{\alpha'}.
\end{eqnarray}
In the single species case, we get $k_{B}T_{c}=(N-1)Gm^2/4$ and $k_{B}T_{c}=-(N-1)\gamma^2/(8\pi)$  and they only differ from the mean field critical temperature by a term of order $1/N$ that tends to zero as $N\rightarrow +\infty$ (recall that the mean field approximation becomes exact for $N\rightarrow +\infty$). The mean field Fokker-Planck equations Eqs. (\ref{e1})-(\ref{e2}) and  Eqs. (\ref{e31})-(\ref{e32}) with rotation have never been studied so far and the present preliminary results (that have been obtained without solving them explicitly) show that they have a very rich behavior. This can stimulate further research from physicists and applied mathematicians.

\end{document}